\documentclass[useAMS,usenatbib]{mnras}
\usepackage{times}
\usepackage{graphicx}
\usepackage{amsmath}
\usepackage{amssymb}
\graphicspath{{./image/}}
\usepackage{lmodern}
\usepackage{hyperref}
\usepackage{makecell}

\newcommand{\HI}{\rm H~{\sc i }}
\newcommand{\HII}{\rm H~{\sc ii }}

\newcommand{\TB}{\delta T_{\rm b}}
\newcommand{\MSUN}{{\rm M}_{\odot}}

\newcommand{\XHII}{x_{\rm HII}}
\newcommand{\TS}{T_{\rm S}}
\newcommand{\TK}{T_{\rm K}}
\newcommand{\TCMB}{T_{\gamma}}
\newcommand{\lya}{\rm {Ly{\alpha}}}
\newcommand{\OmegaB}{\Omega_{\rm B}}
\newcommand{\Omegam}{\Omega_{\rm m}}

\title[Comparison between $\rm C^2$-{\sc ray} and {\sc grizzly}]{Prediction of the 21-cm signal from reionization: comparison between 3D and 1D radiative transfer schemes}

\author[Ghara et.al]{
\parbox[t]{\textwidth}{
Raghunath Ghara$^1$\thanks{Email: ghara.raghunath@gmail.com},
Garrelt Mellema$^1$, Sambit K. Giri$^1$, 
T. Roy Choudhury$^2$, \newline
Kanan K. Datta$^{3}$ $\&$ Suman Majumdar$^{4}$ } 
\vspace*{6pt} \\
$^1$ Department of Astronomy \& Oskar Klein Centre, AlbaNova, Stockholm University, SE-106 91 Stockholm, Sweden\\
$^2$ National Centre for Radio Astrophysics, TIFR, Post Bag 3, Ganeshkhind, Pune 411007, India\\ 
$^3$ Department of Physics, Presidency University, 86/1 College Street, Kolkata - 700073, India\\
$^4$ Department of Physics, Blackett Laboratory, Imperial College, London SW7 2AZ, UK }

\date{Accepted ?; Received ??; in original form ???}

\pubyear{?}

\begin{document}
\label{firstpage}
\pagerange{\pageref{firstpage}--\pageref{lastpage}}
\maketitle


\begin{abstract}
Three-dimensional radiative transfer simulations of the epoch of reionization can produce realistic results, but  are computationally expensive. On the other hand, simulations relying on one-dimensional radiative transfer solutions are faster but limited in accuracy due to their more approximate nature. Here, we compare the performance of the reionization simulation codes {\sc grizzly} and $\rm C^2$-{\sc ray} which use 1D and 3D radiative transfer schemes respectively.  The comparison is performed using the same cosmological density fields, halo catalogues and source properties.  We find that the ionization maps, as well as the 21-cm signal maps from these two simulations are very similar even for complex scenarios which include thermal feedback on low mass halos. The comparison between the schemes in terms of the statistical quantities such as the power spectrum of the brightness temperature fluctuation agree with each other within 10\% error throughout the entire reionization history. {\sc grizzly} seems to perform slightly better than the semi-numerical approaches considered in \citet{2014arXiv1403.0941M} which are based on the excursion set principle. We argue that {\sc grizzly} can be efficiently used for exploring parameter space, establishing observations strategies and estimating parameters from 21-cm observations. 
\end{abstract}



\section{Introduction}
\label{intro}
The Epoch of reionization (EoR), when the primordial neutral hydrogen (\HI) in the intergalactic medium (IGM) was ionized by the radiation from the first sources, is one of the milestone events in the evolution of the Universe. However, many important facts such as the exact timing of this event, the sources responsible for reionization, the impact of the EoR on the later stages of the structure formation, etc.\ are currently not or poorly known. Various probes of the EoR such as observations of high-redshift quasars \citep{becker01, fan2006, Goto11, 2011Natur.474..616M, 2015MNRAS.447..499M}, the cosmic microwave background radiation (CMB) \citep{Zaldarriaga97, Komatsu2011, 2016A&A...594A..15P, 2016A&A...594A..13P}, high-redshift galaxies \citep{Kashikawa11, Bouwens15, Mitra15} and $\lya$ emitters \citep{2007MNRAS.381...75M, 2011MNRAS.414.2139D, 2015MNRAS.452..261C} have been used to constrain some of these unknown facts. However, these probes can only provide limited information. For example, combining the observations of high-redshift quasars and the CMB, one can constrain the probable period of the event to lie between redshifts 15 and 6 \citep{fan2006, Malhotra06, Choudhury06a, mitra2011, mitra2012}.  

The evolving redshifted 21-cm signal from the \HI ~in the IGM has the potential to revolutionize our understanding of reionization \citep{Furlanetto2006, Morales10, Pritchard12}. Therefore a range of low frequency radio telescopes such as the Giant Metrewave Radio Telescope (GMRT)\footnote{http://www.gmrt.tifr.res.in}\citep{ghosh12, paciga13}, the Precision Array for Probing the Epoch of Reionization (PAPER)\footnote{http://eor.berkeley.edu/} \citep{parsons13}, the Murchison Widefield Array (MWA)\footnote{http://www.mwatelescope.org/} \citep{bowman13, tingay13} and the Low Frequency Array (LOFAR)\footnote{http://www.lofar.org/} \citep{van13, 2017ApJ...838...65P} are attempting to detect this faint signal from the EoR. The primary aim of these first generation experiments is to detect the signal in terms of statistical quantities such as the variance, skewness, power spectrum, etc. However, future radio telescopes such as the Square Kilometre Array (SKA)\footnote{http://www.skatelescope.org/} should have enough sensitivity to produce image cubes of the signal \citep{2015aska.confE..10M, ghara16}.

A large number of theoretical models based on either analytic calculations \citep[e.g.,][]{furlanetto04, 2014MNRAS.442.1470P}, semi-numerical simulations \citep{zahn2007, mesinger07, santos08, choudhury09}, or numerical simulations with radiative transfer \citep{Iliev2006,  mellema06, McQuinn2007, shin2008, baek09, Thom09, ghara15a} have been used to study the redshifted 21-cm signal from the EoR. Through these works a fair amount of theoretical understanding of the impact of various physical processes on the signal has been achieved. Aspects which have been considered include the impact of different kind of sources to the redshifted 21-cm signal, thermal feedback and peculiar velocities of the gas in the IGM. The simulations are not only important to extract information from the observations, but also to design observation strategies.

One thing to keep in mind is that the 21-cm observations do not provide direct information about the properties of the sources responsible for ionizing the IGM, minimum mass of the dark matter halos forming stars, thermal feedback, the effect of peculiar velocities of the gas in the IGM, etc. Due to the complexity of the astrophysical effects, one needs to use simulations to extract this information from the observations. In principle, this can be done using parameter estimation techniques such as the Markov chain Monte Carlo (MCMC) \citep{2015MNRAS.449.4246G} or artificial neural networks \citep{2017MNRAS.468.3869S} in which the observations expressed in quantities such as power spectrum, variance or skewness will be compared to the outputs of the simulations. 

The various simulation methods each have their own advantages and drawbacks and when using their results to design observation strategies or to interpret the observational data, one needs to keep in mind the various assumptions made by the models. While simulations of the 21-cm signal from the EoR using 3D radiative transfer methods such as \citet{Iliev2006,  mellema06} are assumed to be the most accurate ones, these require supercomputer facilities to run. On the other hand, simulations performed with semi-numerical methods are orders of magnitudes faster, the simplifying assumptions needed for this may lead to accuracy issues. Thus, it is absolutely necessary to compare the performance of different simulation techniques with each other so as to identify the advantages and drawbacks of these methods. This will be useful for improving the methods and for correctly interpreting the observations.

Studies such as \citet{2011MNRAS.414..727Z} compared a reionization simulation obtained using a full 3D radiative transfer technique with the result from a semi-numerical one.  It showed excellent agreement between these two. Similarly, \citet{2014arXiv1403.0941M} compared semi-numerical simulations and conditional Press-Schechter models with a full 3D radiative transfer simulation performed with $\rm C^2$-{\sc ray} \citep{mellema06}. That study showed that the semi-numerical simulations such as \citet{zahn2007, mesinger07, choudhury09, 2010MNRAS.406.2421S}, which use an excursion-set formalism to create the ionized bubbles, produce an ionization history very similar to the $\rm C^2$-{\sc ray} provided that the halo lists from an $N$-body simulation are used. However, if the halos are found from a conditional Press-Schechter approach, as is for example done in \citet{2009ApJ...703L.167A, mesinger2011}, the fit with the $\rm C^2$-{\sc ray} results is considerably worse.

One-dimensional radiative transfer simulations such as {\sc bears} \citep{Thom09}, \citet{ghara15a} (hereafter {\sc grizzly}) assume that the effect of each source is perfectly spherical and use special recipes to deal with overlapping ionized and/or heated regions. Due to these simplifying assumptions they are fast but can also be expected to be less accurate than full 3D radiative transfer schemes. Previously, \citet{Thom09} performed a limited comparison between the 1D radiative transfer code {\sc bears} and the 3D radiative transfer code {\sc crash} \citep{2001MNRAS.324..381C} for a simulation volume of size 12.5 $h^{-1}$ comoving megaparsec. However, before considering these 1D schemes for parameter estimations \citep[e.g.,][]{patil2014}, their performance for larger simulation volumes and larger numbers of sources is required.

In this paper we present a detailed comparison between the performance of the 1D radiative transfer code {\sc grizzly} \citep{ghara15a} with the 3D radiative transfer code $\rm C^2$-{\sc ray}. We use the same initial conditions such as the density fields, halo catalogues, source models, and compare the reionization history produced by these two schemes. We consider bubble size distributions as well as cross-correlations between ionization maps, while for the redshifted 21-cm signal we focus on statistical quantities such as the evolution of the power spectra. We consider two different models: ($i$) reionization with only large mass halos which do not suffer from thermal feedback, ($ii$) reionization with both high and low mass halos where the latter are assumed to be sensitive to thermal feedback. Case ($i$) is identical to the one used in \citet{2014arXiv1403.0941M} to compare semi-numerical and full radiative transfer results which therefore allows the comparison between semi-numerical, 1D RT and 3D RT results.

Our paper is structured in the following way. In section \ref{simu}, we describe the simulations used in this paper. In particular different subsections describe the $N$-body simulation and the source model used in this study as well as a brief description of the underlying algorithms of $\rm C^2$-{\sc ray} and {\sc grizzly}. In section \ref{res}, we present our results before we conclude in section \ref{conc}. Throughout the paper, we have chosen the cosmological parameters   $\Omegam=0.27$, $\Omega_\Lambda=0.73$, $\OmegaB=0.044$, $h=0.7$ consistent with the $Wilkinson ~Microwave ~Anisotropy ~Probe$ ($WMAP$) results \citep{2013ApJS..208...19H} and within the error bars consistent with $Planck$ \citep{Planck2013}.


\section{SIMULATIONS}
\label{simu}
Below we discuss the radiative transfer methods along with the source model and $N$-body simulation used in this study.

\subsection{$N$-body simulation}
\label{n_body}
Both radiative transfer methods considered in this study use the results of the same $N$-body simulation. This dark matter only $N$-body simulation was carried out using the publicly available code {\sc cubep}$^3${\sc m}\footnote{\tt http://wiki.cita.utoronto.ca/mediawiki/index.php/CubePM} \citep{Harnois12}. The details of the simulations are: (i) the size of the simulation cube is 163 comoving megaparsec (cMpc), (ii) the number of dark matter particles is $3072^3$, (iii) the fine grid used by the particle-particle-mesh method is $6144^3$, (iv) the mass  of the dark matter particles is $5.47\times 10^6 ~\MSUN$. The details of the simulation can also be found in \citet{2012MNRAS.423.2222I}. Although this $N$-body simulation is by now rather old, we use it to enable direct comparison to the results in \citet{2014arXiv1403.0941M} who compared the performance of a semi-numerical method to $\rm C^2$-{\sc ray}.

Snapshots of the density field are generated from redshift 20.134 to redshift 6 using equal time intervals of 11.5 megayears (Myr). The resolution of these gridded density fields is $256^3$. {\sc cubep}$^3${\sc m} also produces dark matter halo lists using an on the fly halo finder based on the spherical over-density method. The lowest mass halos produced have a mass of $\sim 10^8 ~\MSUN$ which corresponds approximately to the mass needed to initiate star formation driven by atomic cooling.

\subsection{Source model}
\label{source_rt}
Our knowledge about the sources of reionization is very limited. Different efforts, both observational and modelling, have been undertaken to improve our understanding of the impact of various sources such as primordial galaxies \citep{2014MNRAS.442.2560W,  2016ApJ...833...84X}, mini-quasars \citep{2003ApJ...596...34B, 2007MNRAS.375.1269Z, 2009ApJ...701L.133A, 2012MNRAS.425.2974T} and high-mass X-ray binaries  \citep{ Fialkov14, 2014MNRAS.440L..26K, 2014MNRAS.445.2034K} on the expected 21-cm signal from the EoR. However, the individual contributions of these sources towards reionization remain quite uncertain.

In this study, we assume the sources of ionizing radiation formed in the dark matter halos. We further assume their spectral energy distribution can be approximated by a blackbody of an effective temperature of 50,000 K, which is appropriate for a massive population II star. We normalize the spectrum such that the rate of production of the ionizing photons is
\begin{equation}
\dot{N_{\gamma}}=g_{\gamma}\frac{M_{\rm h} \Omega_{\rm B}}{(10 ~\rm Myr)\Omega_m m_{\rm p}},
\label{equ_ngamma}
\end{equation}
where $g_{\gamma}$ is the source ionization efficiency coefficient, $M_{\rm h}$ and $m_{\rm p}$ are the mass of the dark matter halo and proton mass respectively. For the fiducial model considered here, we choose $g_{\gamma}=21.7$ for the sources formed in dark matter halos with mass $M_{\rm h} \geq 2.2 \times 10^9 ~\MSUN$, which will produce $\dot{N_{\gamma}}=1.33 \times 10^{43} \times \frac{M_{\rm h}}{\MSUN} ~\rm s^{-1}$. This is a similar choice as the $L3$ model in \citet{2012MNRAS.423.2222I}.

In addition, we consider a more complex reionization scenario where we use all halos of mass $M_{\rm h} \geq 10^8 ~\MSUN$. We divide these in two populations, low mass atomically cooling halos (LMACHs) in the mass range $10^8 \leq M_{\rm h} <10^9~\MSUN$ and high mass atomically cooling halos (HMACHs) in the mass range $M_{\rm h} \geq 10^9 ~\MSUN$. The LMACHs are assumed to be sensitive to thermal feedback and stop producing ionizing photons once they are located in an cell which is more than 10 percent ionized. The HMACHs are taken to be insensitive to thermal feedback. While still active the LMACHs produce ionizing photons with an efficiency $g_\gamma = 130$, whereas the HMACHs have $g_\gamma = 8.7$. The higher efficiency of the LMACHs is motivated by either a larger contribution from Pop~III stars or a higher fraction of ionizing photons escaping. In the $\rm C^2$-{\sc ray} results these parameters result in a rather early end of reionization around $z=8.5$ and a Thomson optical depth value of 0.08, consistent with the value determined by WMAP and within the 2$\sigma$ range of Planck. This scenario is identical to the $L1$ model of \citet{2012MNRAS.423.2222I}. The details of these scenarios are given in Table \ref{tab1}.

\begin{table*}
\centering
\small
\tabcolsep 6pt
\renewcommand\arraystretch{1.5}
   \begin{tabular}{c c c c c c}
\hline
\hline
    Model  & Min. halo ($\MSUN$)  & Thermal feedback & \makecell{$g_{\gamma}$ \\ LMACH}	& \makecell{$g_{\gamma}$ \\ HMACH} & \makecell{Model used in \\  \citet{2012MNRAS.423.2222I}} \\

\hline
\hline
    MODEL I (fiducial)      & $2.2\times 10^9$        & No   	    &	0 & 21.7 & $L3$			\\
    MODEL II      & $10^8$        & yes   	    &	130 & 8.7 & $L1$			\\
\hline
\end{tabular}
\caption[]{Details of the scenarios considered in this paper. LMACH and HMACH represent the low-mass and high-mass atomically cooling halo respectively.}
\label{tab1}
\end{table*}


\subsection{3D radiative transfer simulation $\rm C^2$-{\sc ray}}
\label{3d_rt}
The 3D radiative transfer method considered in this work uses `Conservative Causal Ray-tracing method' ($\rm C^2$-{\sc ray}). The details of the method can be found in \citet{mellema06}. The method is briefly described in the following steps.
\begin{itemize}
\item It starts with preparing the source list in a random order at each redshift.

\item Given the SED of the sources, the total photo-ionization rate ($\Gamma$) is calculated at time $t$ at each cell in the simulation box including contributions from all the sources. This step takes into account the time evolution of the neutral fraction of the cells during the time step which affects the optical depth from the sources.

\item The history of the ionization fraction of hydrogen ($\XHII$) is estimated by solving the non-equilibrium equation,

\begin{equation}
\frac{{\rm d}\XHII}{{\rm d}t}=\left(1-\XHII\right)\left(\Gamma + n_eC_{\rm H}\right)-xn_eC\alpha_{\rm H},
\label{3rt_equ}
\end{equation}
where $n_e$ is the electron density at the cell, $C_{\rm H}$ and $\alpha_{\rm H}$ represent the collisional ionization and recombination coefficients for hydrogen respectively. The quantity $C$ is the clumping factor (which can be defined as $\left<n^2\right>/\left<n\right>^2$, where $n$ is the baryon number density) which accounts the clumpiness of the IGM. Here we have taken $C=1$ but in general it can be larger than 1 or even position dependent.
\end{itemize}

$\rm C^2$-{\sc ray} was compared to other fully numerical radiative transfer codes in \citet{2006MNRAS.371.1057I} and \citet{2009MNRAS.400.1283I} and was found to produce accurate results. In this study we do not consider the temperature evolution of the intergalactic gas, only its ionized state.


\subsection{1D radiative transfer simulation {\sc grizzly}}
\label{1d_rt}
The details of the one-dimensional radiative transfer method can be found in \citet{ghara15a}. The method mainly follows the {\sc bears} algorithm  \citep{Thom09} with a few changes compared to the original method. To express its relation to the original {\sc bears} method, we have chosen to call it {\sc grizzly}. The basic idea of both methods is not to solve the one-dimensional radiative transfer equations on the fly while generating the ionization maps. Rather, they use pre-generated ionization profiles to construct the ionization maps. Here, we briefly summarize the most important steps of the algorithm.

\begin{itemize}
\item We start by choosing the range for the key source parameters: redshift, density contrast of the uniform background IGM, luminosity of the source, age of the source. We then generate a large number of ionization and kinetic temperature profiles around the sources for different combinations of these parameter values. Next we create a list of radii of the \HII ~bubbles as a function of different parameters. We assume the radius of the \HII ~region to be the radius at which the ionization fraction drops to 0.5 from the location of the source. This library of 1D profiles only needs to be compiled once for a given cosmology and spectral energy distribution of sources.

\item For a given redshift and a certain halo position, we determine the luminosity and the age of the halo according to a source recipe, see Sect.~\ref{sec:grizzlysource}. From this we determine the size of the \HII ~bubble around it using the density field and the pre-generated list of the \HII ~bubble radii. This step is done iteratively. We start with a small radius around the centre of the source, calculate the spherically averaged over-density of the IGM within this radius and then estimate the radius of \HII ~bubble from the list. We change the initial choice of radius such that the estimated radius of the \HII ~region matches with the chosen radius for the same over-density. The parameter values corresponding to that radius are then associated with the source and the corresponding profiles will be used to generate the ionization and kinetic temperature maps.

\item The procedure described in the previous step is performed for all halos at the redshift being considered. Although the number of overlaps between the individual \HII ~regions around the halos are negligible during the initial stages of reionization, they become important at the later stages of reionization. To incorporate the effect of overlap, we estimate the unused ionizing photons for each overlapping \HII ~region and distribute them around the overlapping regions such that each photon ionizes one hydrogen atom.

\item Next, we use the pre-generated 1D ionization profiles corresponding to each halo to calculate the ionization fraction at the partially ionized region beyond the \HII ~regions. The ionization fraction in the overlapping partially ionized regions is then expressed as
\begin{equation} 
x_{\rm HII}({\mathbf x}) =\frac{\sum_i x^{i}_{\rm HII-1D}({\mathbf x}) \times n^{i}_{\rm H-1D} \times \left(1- x^{i-1}_{\rm HII}({\mathbf x})\right)}{n_{\rm H}({\mathbf x})},
\label{p5_overlapion}
\end{equation}
where the summation symbol represents a sum over all overlapping sources. Here, $x^{i-1}_{\rm HII}({\mathbf x})$ denotes the ionization fraction obtained after considering overlapping \HII ~ regions from $i-1$ number of sources. We take $x^{0}_{\rm HII}({\mathbf x})=0$. The term $\left(1- x^{i-1}_{\rm HII}({\mathbf x})\right)$ takes care of the fact that the $i$th source encounters an already partially ionized IGM, due to overlap between previous $i-1$ number of sources, before contributing to ionization.  

\item Next, we generate the kinetic temperature maps using a  correlation of the neutral fraction and the gas temperature in the partially ionized region \citep[for details, see][]{ghara15a}.  

\end{itemize}

\subsubsection{Implementation of source model}
\label{sec:grizzlysource}
In the model scenarios considered in this paper, we assume that dark matter halos always have ongoing star formation, unless they are suppressed through thermal feedback, see Sect.~\ref{sec:grizzlysuppresion}. To select a one-dimensional profile from the library we need to specify a luminosity and age for the halo. The original {\sc bears} code and \citet{ghara15a} used a fixed age for the halos, typically $\sim$ 10 Myr and also used a luminosity calculated from the instant mass of a halo at the redshift under consideration. Those models assume that the stars inside the galaxies efficiently emit hydrogen ionizing photons until an age of $\sim 10$ Myr.

However, we found this source model to be inconsistent with the one used by $\rm C^2$-{\sc ray} in terms of the cumulative number of ionizing photons being produced. We therefore implemented another source model. In this model we follow the growth of halos and find both the age of the halo as well as the time-averaged mass (to be called the effective mass $M_{\rm eff}$). We then calculate the luminosity based on $M_{\rm eff}$ using Eq.~\ref{equ_ngamma} and use this together with the age to find the appropriate profile from the library. The {\sc grizzly} code thus does not track the history of the state of the IGM gas but rather uses the history of the halos to calculate the state of the IGM at a given redshift.

It is not entirely trivial to calculate the age of the halos accurately from the halo lists as halos can move to new cells or multiple halos can merge. In principle, one needs to track the halos using a merger-tree algorithm. However, as no merger tree is supplied with the halo lists, we have used a simple method to account for their growth. For every pair of redshifts $z_1$ and $z_2$ we first search for a halo at the same location and it is found, we associate the change in mass with the halo whose history we are tracking. If a halo disappears from its previous grid position in some time step, we inspect bordering grid cells. If we find a newly formed heavier halo in a bordering grid cells, we assume that the identified halo must be the present state of the halo which disappeared. If we are not able to find any newly formed halo in the neighbourhood, then we assume that a merger happened. In this case, we find previously existing heavier halo at the neighbour and set its age as the maximum of the ages of the identified and disappeared halo. Although this a posteriori reconstruction of the halo histories is rather crude, it does allow us to obtain the correct cumulative number of ionizing photons produced during reionization.

\subsubsection{Implementation of thermal feedback}
\label{sec:grizzlysuppresion}
In the second $\rm C^2$-{\sc ray} simulation we include LMACHs which are fully suppressed once their environment has been sufficiently ionized. The criterion used is that when the ionization fraction in the cell containing the LMACH is larger than 0.1, thermal feedback is assumed to stop the production of ionizing photons\footnote{The suppression of the low mass halos is determined by the kinetic temperature of the region. For example, one can assume no star formation within halos with masses $< 10^9 ~\MSUN$ if it is formed within a region with $\TK$ larger than $10^4$ K. Following \citet{2012MNRAS.423.2222I} we use $x_{\rm HII}$ as proxy for the temperature and assume that if a region is more ionized than 0.1, it will be hot enough to suppress further star formation.}.

To implement this recipe in {\sc grizzly} it is not trivial as the code only places ionized regions around active sources at the redshift of interest. We implemented the following simple prescription to take into account thermal feedback effects in {\sc grizzly}. We use the ionization map from the previous time step to select the active sources at a certain time step using the condition on the ionization fraction. We add to these the previously active sources which became inactive at the present time step but assign to them their effective mass and age corresponding to the last time they were active. In principle one needs to consider the recombination in the already ionized regions to correctly account for the suppression. However, it is not straightforward to implement recombination in an already ionized region in {\sc grizzly}. This is also true for other semi-numerical simulations such as models considered in \citet{2014arXiv1403.0941M, 2014MNRAS.440.1662S}. However, recombinations play a minor role, especially since previously suppressed LMACH halos become active again after their halos have grown into HMACHs and the amount of time that recombinations act upon the relic \HII ~regions will thus be limited.

Since the criterion for suppression is $x_\mathrm{HII} > 0.1$, the treatment of partially ionized cells is important. We therefore implemented an improvement over the previous version of {\sc grizzly} when estimating the ionization fraction in cells containing an ionization front. We now assign the volume averaged ionized ionization fraction to these cells calculated from the higher resolution 1D profiles.

\subsection{Differences between $\rm C^2$-{\sc ray} and {\sc grizzly}}
\label{sec:differences}
It is important to realise that $\rm C^2$-{\sc ray} and {\sc grizzly} treat the calculation of the reionization history very differently. $\rm C^2$-{\sc ray} follows the time evolution of the ionization fractions by taking the state at the end of the previously calculated time step as initial condition for the new time step. This means that any simulation has to be followed from the emergence of the first sources to a chosen end point, for example when the IGM reaches an average ionization fraction of 99 percent. The advantage is that any time-dependent effects such as recombinations and changes in the source luminosity or position are treated correctly.

The {\sc grizzly} as well as the related {\sc bears} code do not use the previous state of the IGM as an initial condition for the next time step but instead construct the state of the IGM from scratch for every output. The time evolution is captured by considering the growth history of dark matter halos when calculating their luminosity and in the case of suppressed halos by using the recipe described above. This approach has as an advantage that it is not required to calculate the entire reionization history if this is not needed. The disadvantage is that time-dependent effects such as recombinations, radiative cooling and changes in the source properties have to be implemented in an approximate manner. These 1D radiative transfer reionization codes share this property with the semi-numerical methods based upon the excursion set approach such as {\sc 21cmfast} \citep{2011MNRAS.411..955M}.

\begin{figure*}
\begin{center}
\includegraphics[scale=0.47]{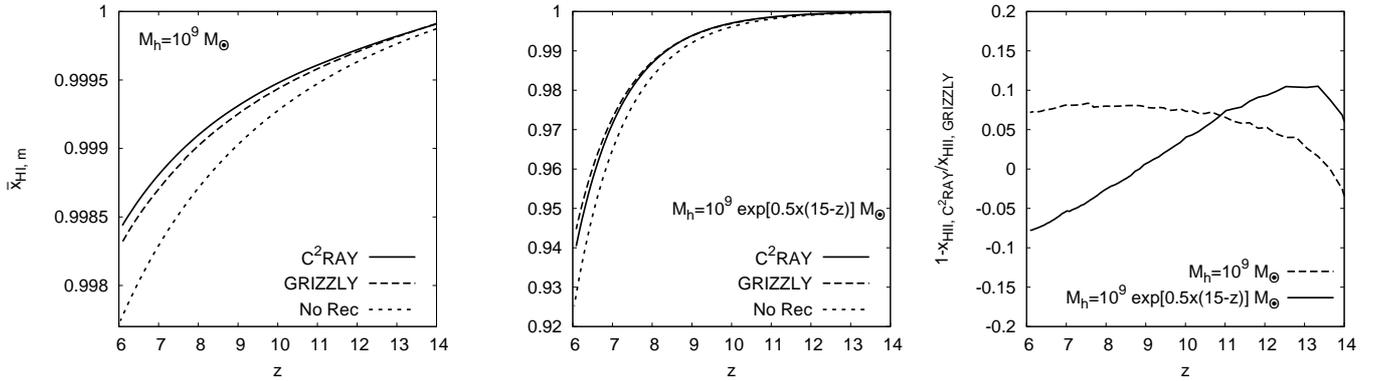}
    \caption{Left-hand panel: Redshift evolution of the mass averaged neutral fraction due to a single source in a simulation volume of size 20 $h^{-1}$ cMpc. The solid and dashed curves represent $\rm C^2$-{\sc ray} and {\sc grizzly} respectively both including recombination processes, while the dotted curve corresponds to {\sc grizzly} without any recombination processes. The emission rate of the ionizing photons is proportional to the mass of the halo which is chosen to be constant ($10^9 ~\MSUN$) in this case. Middle-panel: Same as the left-hand panel, but for a dark matter halo whose mass changes exponentially with redshift. Right-hand panel: Relative difference between the ionization fractions calculated from $\rm C^2$-{\sc ray} and {\sc grizzly}  as a function of redshift for the test scenarios shown in the left and middle panels of the figure.}
   \label{image_p5_ion_single}
\end{center}
\end{figure*}

\subsection{21-cm signal}
\label{res_21cm}
The differential brightness temperature $\TB$ of the 21-cm signal can be expressed as,
\begin{eqnarray}
 \TB (\mathbf{x}, z)  \!\!\!\! & = & \!\!\!\! 27 ~ x_{\rm HI} (\mathbf{x}, z) [1+\delta_{\rm B}(\mathbf{x}, z)] \left(\frac{\OmegaB h^2}{0.023}\right) \nonumber\\
&\times& \!\!\!\!\left(\frac{0.15}{\Omegam h^2}\frac{1+z}{10}\right)^{1/2}  \left(1-\frac{\TCMB}{\TS} \right)\,\rm{mK},
\nonumber \\
\label{brightnessT}
\end{eqnarray}
where the quantities  $x_{\rm HI}$, $\delta_{\rm B}$ and $\TCMB(z)$ = 2.73 $\times (1+z)$ K denote the neutral hydrogen fraction, baryonic density contrast and the CMB brightness temperature, respectively, each at position $\mathbf{x}$ and redshift $z$. $\TS$ represents the spin temperature of hydrogen in the IGM.

In this study we will adopt the high spin temperature approximation: $\TS \gg \TCMB$. In this case $\TB$ becomes insensitive to the actual value of the spin temperature. This approximation is valid when the spin temperature is coupled to the IGM gas temperature through $\lya$ coupling and the IGM has been sufficiently heated ($\TK \gg \TCMB$) by the X-ray sources. Most reionization scenarios reach this state well before substantial ionized regions form \citep{2017MNRAS.472.1915C}.

After generating the ionization fraction cubes as described in sections \ref{3d_rt} and \ref{1d_rt}, it is straightforward to generate the differential brightness temperature cubes using Eq.~\ref{brightnessT}. Below we will compare $x_{\rm HII}$ and $\TB$ results at fixed redshifts. We will not consider the effect of peculiar velocities in the IGM (leading to the so-called redshift-space distortions). The comparison of fully numerical and semi-numerical models in \citet{2014arXiv1403.0941M} suggest that redshift-space distortions will not change the main conclusions of the comparison.


\section{Results }
\label{res}

In this section, we compare the different outputs from the two simulation codes considered in this work, namely $\rm C^2$-{\sc ray} and {\sc grizzly}. Before evaluating the different scenarios listed in Table \ref{tab1}, we first consider two single source scenarios to test the impact of the fundamental difference between the two codes as described in Sect.~\ref{sec:differences}. 

\subsection{Test scenarios: single sources}
\label{app_test}

To gauge the basic differences between $\rm C^2$-{\sc ray} and {\sc grizzly} we follow the growth of an ionized region for two single source scenarios. For both we take the IGM around the source to have a uniform density at the mean value of the Universe. We choose the size of the simulation box to be 20 $h^{-1}$ cMpc and use a grid of size $100^3$. We assume that the source formed at redshift 15 and we follow the impact of the source on the IGM up to redshift 6 using time steps of 10 Myr.

\begin{itemize}
\item TEST A: The source is hosted by a dark matter halo with constant mass which we take to be $10^9 ~\MSUN$. We choose an efficiency factor $g_{\gamma} = 21.7$, as defined in Equation \ref{equ_ngamma}, which sets the rate of production of ionizing photons to $\dot{N_\gamma} = 1.33\times 10^{52} ~\rm s^{-1}$. 

\item TEST B: The source is hosted by a dark matter halo with a mass which changes exponentially with redshift as \citep[see, e.g,][]{2002ApJ...568...52W},
\begin{equation}
M_h(z)=m_i \times \exp\left[\beta (z_i -z)\right],
\end{equation}
where $m_i=10^9 ~\MSUN$, $z_i=15$, $\beta = 0.5$. This scenario mimicks the typical growth of dark matter halos and thus will show the impact of capturing source evolution through an effective mass $M_\mathrm{eff}$ in {\sc grizzly}. We choose the same efficiency $g_{\gamma}$ as in Test A. 
\end{itemize}

The performance of the methods for these test scenarios is shown in Figure \ref{image_p5_ion_single}. The left hand and middle panel shows the evolution of the neutral fraction in the simulation volume for TEST A and B respectively, whereas the right hand panel shows relative difference between the two codes. We see that they very similar results for both tests with differences less than 10 per cent. However, for TEST A the ionization front grows faster in the {\sc grizzly} results. The probable reason is that {\sc grizzly} underestimates the impact of recombinations. While placing the \HII ~bubble around the sources, we assume that the source has been emitting ionizing photons at a constant rate determined by its effective mass ($M_{\rm eff}$) during its entire age and we include the effect of recombinations using the densities at the redshift which we are considering. Since the expansion of the Universe causes densities to be higher at higher redshifts, this will underestimate the impact of recombinations over the age of the source. $\rm C^2$-{\sc ray} calculates the recombination rates at every time step using the densities at that redshift and thus captures this evolution of the recombination rate.

Interestingly, the match between the two codes is better for TEST B where the halo mass is increasing in time (see the middle panel of Figure \ref{image_p5_ion_single}). Since in this case most photons are emitted at later times and the growth of the \HII ~ region is faster at later times, the effect of underestimating recombinations is less severe in this case. Below when considering more complex simulations with multiple sources we will see further demonstrations of the impact of recombinations. 

We conclude that {\sc grizzly} is capable of approximating the expansion of an ionization front around a growing halo to within 10 per cent accuracy, giving confidence that the effective mass approach works well. When considering the multiple sources case below, larger differences than the ones found here will have to be due to how {\sc grizzly} deals with overlapping \HII ~regions and source suppression.

\begin{figure}
\begin{center}
\includegraphics[scale=0.92]{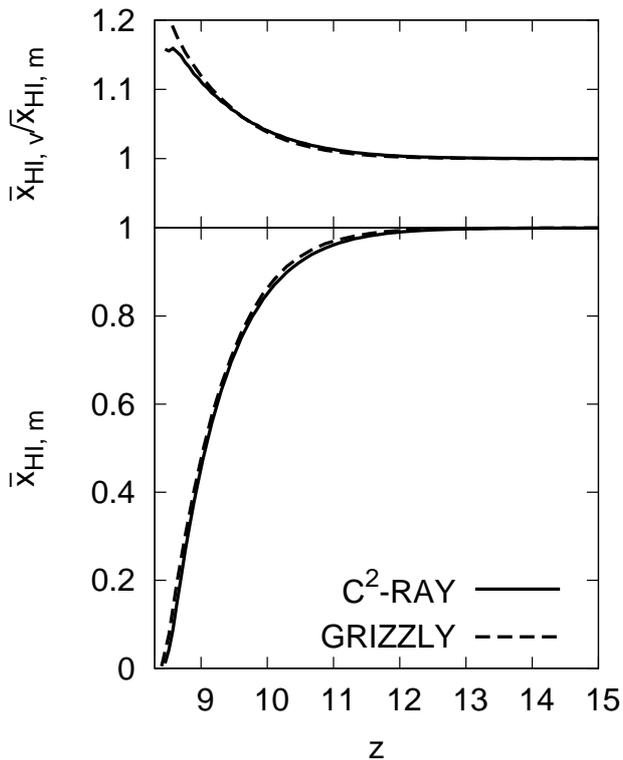}
    \caption{Bottom panel: Redshift evolution of the mass averaged neutral fraction for Model I from $\rm C^2$-{\sc ray} (solid curve) and {\sc grizzly} (dashed curve). Top panel: Same but showing the redshift evolution of the ratio of the volume and mass averaged neutral fractions. }
   \label{image_p5ionfrac}
\end{center}
\end{figure}

\begin{figure}
\begin{center}
\includegraphics[scale=0.84]{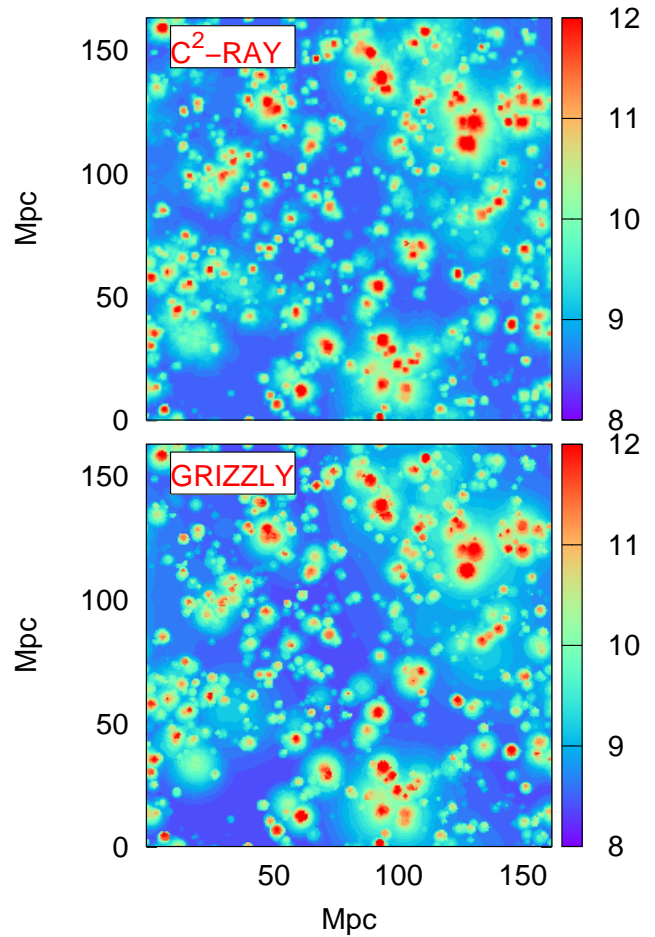}
    \caption{Top panel: Image showing the redshifts of reionization for pixels in a slice from the simulation volume for Model I generated using the results from $\rm C^2$-{\sc ray}. Bottom panel: Same as the top panel but generated from the {\sc grizzly} results.}
   \label{image_p5_zreion}
\end{center}
\end{figure}

\begin{figure}
\begin{center}
\includegraphics[scale=0.8]{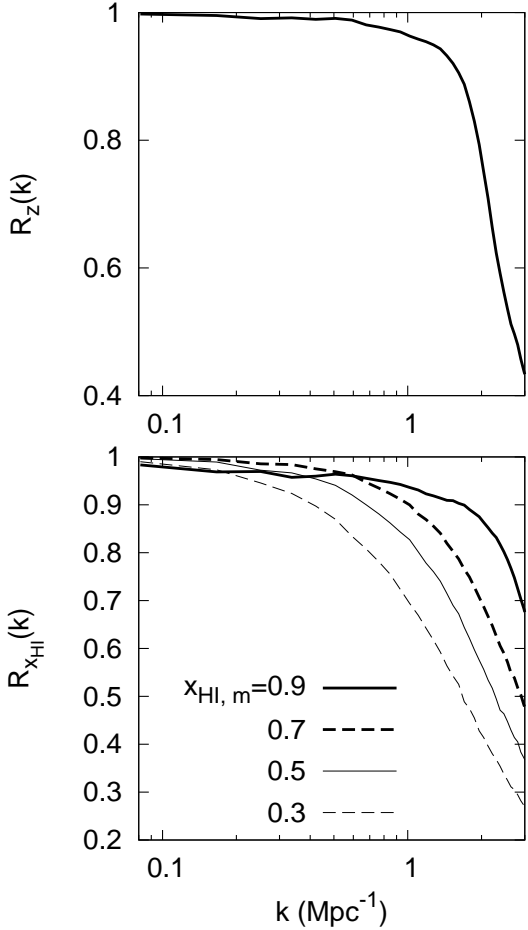}
    \caption{Top panel: Scale dependence of the cross-correlation coefficient of the redshift of reionization maps from $\rm C^2$-{\sc ray} and {\sc grizzly} simulations for Model I. Bottom panel: Same as the top panel but for the neutral fraction maps at different stages of reionization.}
   \label{image_p5_crosstbxhi}
\end{center}
\end{figure}

\begin{figure}
\begin{center}
\includegraphics[scale=0.84]{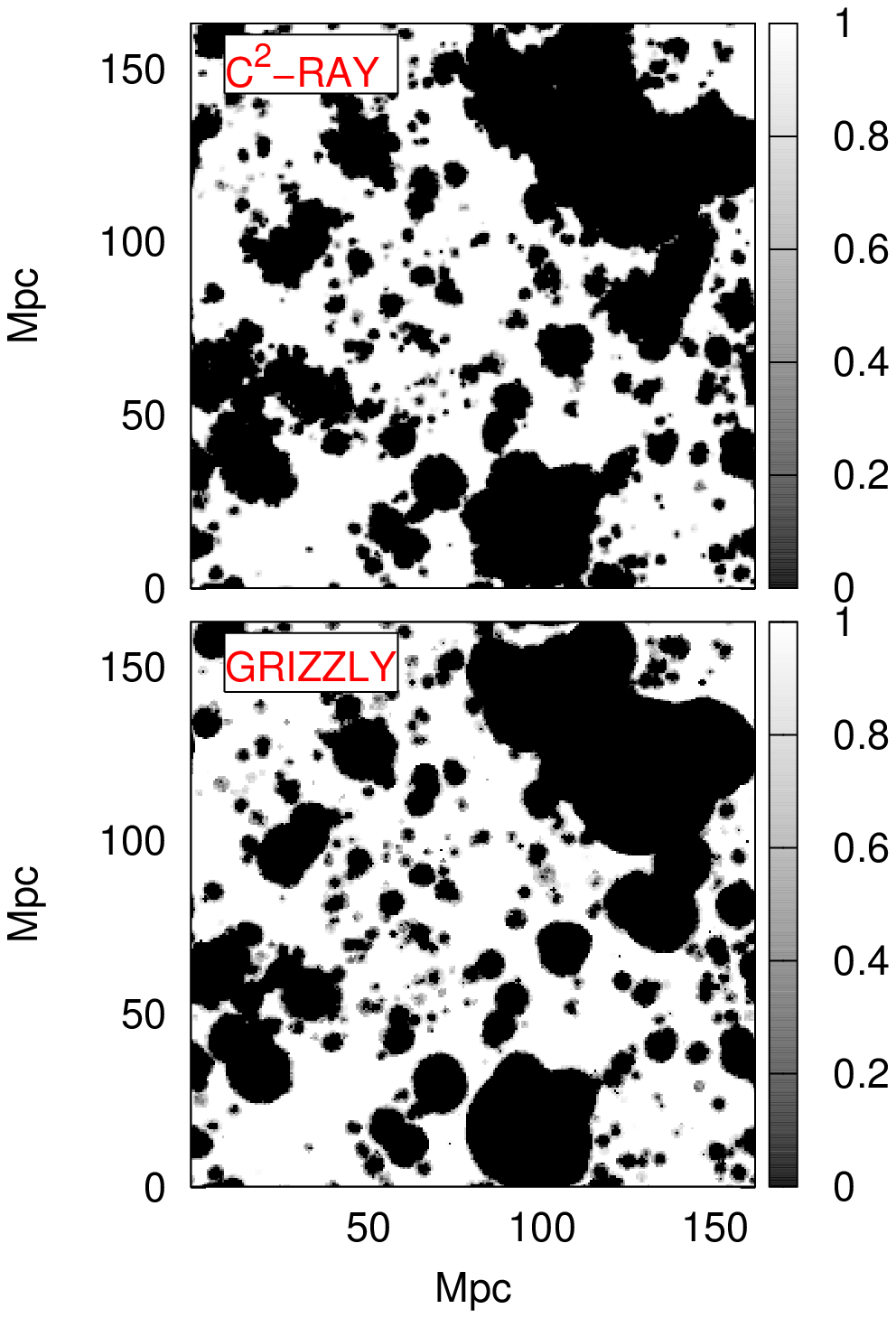}
    \caption{Images of the neutral fraction for the same slice through the simulation box as in Fig.~\ref{image_p5_zreion}. Black is ionized, white is neutral. The top and bottom panels show the results from $\rm C^2$-{\sc ray} and {\sc grizzly} respectively. The results are for Model I, $z=9.026$ when the mass averaged neutral fraction is $\sim$ 0.5.  }
   \label{image_p5_mapxhizmid}
\end{center}
\end{figure}

\subsection{Model I : Large halos only}
\label{sec_model2}
Now, we will consider a model where the reionization is driven by halos with mass $M_{\rm h} \geq 2.2 \times 10^9 ~\MSUN$ in a volume 163 comoving Mpc across. These halos are never suppressed. This model is same as the $L3$ model in \citet{2012MNRAS.423.2222I} and was also used in \citet{2014arXiv1403.0941M} to compare semi-numerical simulation techniques with a fully numerical one.

Not taking into account the time needed to construct the library of 1D ionization profiles, the simulation run time for the whole reionization history is  $\sim 5.5$ CPU hours\footnote{As mentioned above, {\sc grizzly} does not track the ionization history to create the ionization map at a certain redshift for simple models where thermal feedback is ignored. Thus, the run time will be much shorter if only a single snapshot of the ionization fields is needed. For example, {\sc grizzly} generates the neutral fraction map at redshift 9.026 with mass averaged neutral fraction $\sim 0.5$ in $\sim 10$ CPU minutes. In presence of thermal feedback, the estimated time is higher which is also true for other codes such as the 21cmFAST \citep[see, e.g.,][]{2014MNRAS.440.1662S}.} for {\sc grizzly}. This number is without generating the $\TK$ maps and $\lya$ flux maps which are not being used here. For comparison, the entire $\rm C^2$-{\sc ray} simulation took $\sim 10^5 $ CPU hours. So there is a factor $\sim 10^5$ gain in computing time when using {\sc grizzly}.

\subsubsection{Reionization history}
\label{RES_his}
Using the source model described in section \ref{source_rt} and following the methods described in sections \ref{3d_rt} and \ref{1d_rt}, we generate the ionization maps in the simulation box. The ionization histories from these two simulations are shown in the bottom panel of Figure \ref{image_p5ionfrac}. In this scenario, the first sources appear at redshift 19 and ends around $z \sim 8.4$. The Thomson scattering optical depth from the {\sc grizzly} and $\rm C^2$-{\sc ray} simulations are 0.0696 and 0.0697 respectively. As can be seen from the figure, the evolution of the mass averaged neutral fraction as a function of redshift from these two simulations are very similar.

The top panel of Figure \ref{image_p5ionfrac} shows the redshift evolution of the ratio of the volume and mass averaged neutral fraction for this reionization history. As the reionization scenario considered here is `inside-out', high-density regions are ionized first. This leads to a higher volume averaged neutral fraction than the mass averaged neutral fraction. Also for this ratio we find very good agreement between the two codes. This suggest that the ionization maps from these two schemes will be very similar. The small differences may originate from the approximations that go into the implementation of the ionization and differences in the treatment of recombinations in the two schemes. This agreement is better than for any of the semi-numerical results from \citet{2014arXiv1403.0941M}, see figure~1 in that paper.

To illustrate the similarity between the reionization histories from these two methods even further, we show the map of redshifts of reionization of each pixel in a chosen slice from the simulation boxes in Figure \ref{image_p5_zreion}. We define the redshift of reionization of a region to be the redshift when the ionization fraction first reaches 0.5. The maps in Figure \ref{image_p5_zreion} from {\sc grizzly} and $\rm C^2$-{\sc ray} are visually quite similar. The same slice for the semi-numerical methods is shown in Figure~2 of \citet{2014arXiv1403.0941M} and shows larger differences. To quantify the similarity of these two maps, we use the Pearson cross-correlation coefficient defined as,
\begin{equation}
\chi_{ab}=\frac{\sum_{i}{(a_i-\bar{a})(b_i-\bar{b})}}{\sqrt{\sum_{i}(a_i-\bar{a})^2}\sqrt{\sum_{i}(b_i-\bar{b})^2}},
\label{eq_pearson}
\end{equation}
where $\bar{a}$ and $\bar{b}$ represent the mean of the maps $a$ and $b$ respectively and $i$ is the pixel index. The Pearson-cross-correlation coefficient calculated for the maps at the top and bottom panels of Figure \ref{image_p5_zreion} is 0.97.

To compare two different maps, we define the cross-correlation coefficient of a quantity $Q$ ($R^{AB}_Q$) as given below.
\begin{equation}
R^{AB}_{Q} (k) = \frac{P_{AB}(k)}{\sqrt{P_{AA}(k)P_{BB}(k)}},
\label{equ_cross}
\end{equation}
where  $P_{AA}$ and $P_{BB}$ are the auto power spectrum of the fields  $A$ and $B$ respectively and the $P_{AB}$ is the cross power spectrum between them. Here, we will consider $A$ and $B$ as {\sc grizzly} and $\rm C^2$-{\sc ray} respectively. The quantity $Q$ could be the redshift of reionization, neutral fraction, brightness temperatures, etc.

The cross-correlation coefficient for the redshift of reionization cubes $R_z$ is found to be larger than 0.95 for scales with $k \lesssim 1 ~\rm Mpc^{-1}$ (see top panel of Figure \ref{image_p5_crosstbxhi}). This suggests that the reionization histories from these two simulations are very similar at large scales. Although the correlation drops at scales correspond to  $k \gtrsim 1 ~\rm Mpc^{-1}$, it is clearly still better than the semi-numerical schemes considered in \citet{2014arXiv1403.0941M} (see Figure 3 of their paper).


\subsubsection{Morphology of \HII\ regions}
\label{sec_ion_map}
Next we compare the morphology of the ionized regions from the two simulations using maps of the neutral fraction. We show the same slice from the neutral fraction cubes from $\rm C^2$-{\sc ray} and {\sc grizzly} in the two panels of Figure \ref{image_p5_mapxhizmid}. Visually, the maps are quite similar. The Pearson cross-correlation coefficient calculated for these two slices is 0.93. One can notice that the ionized bubbles in the slice from {\sc grizzly} simulation appear to be more spherical than those from $\rm C^2$-{\sc ray}. This is due to the application of 1D radiative transfer which erases the directional dependence of the \HII ~bubbles. However, the shapes of the bubbles become irregular once there is a significant amount of overlap between the \HII\ regions.  

The scale dependent cross-correlation coefficients for maps of the neutral fraction at different stages of reionization from {\sc grizzly} and $\rm C^2$-{\sc ray} are shown in the bottom panel of Figure \ref{image_p5_crosstbxhi}. One can see very high correlation at large length scales, which are the ones of prime interest to the first generations of radio telescopes. We find that the cross-correlation coefficient of the neutral maps remains very close to unity (within $\sim 5$ per cent) at scales $k \leq 1 ~\rm Mpc^{-1}$ until the universe is 50\% ionized. There is a consistent shift to smaller $k$ values in the scale at which the correlation coefficient reaches 0.9 as reionization progresses.  Although the correlation of the ionization fields drops for small length scales, the values actually remain higher than those for the semi-numerical models considered in \citet{2014arXiv1403.0941M}. For example, the cross-correlation coefficients of the neutral fractions between  $\rm C^2$-{\sc ray} and different semi-numerical schemes were found to be smaller than $\sim 0.4$ at a scale $k \sim 2 ~\rm Mpc^{-1}$ at a stage with neutral fraction $x_{\rm HI} \sim 0.5$, while the corresponding value for {\sc grizzly} is $\sim 0.6$.

\subsubsection{Size distribution of \HII\ regions}
\label{res_bubble}
\begin{figure}
\begin{center}
\includegraphics[scale=0.8]{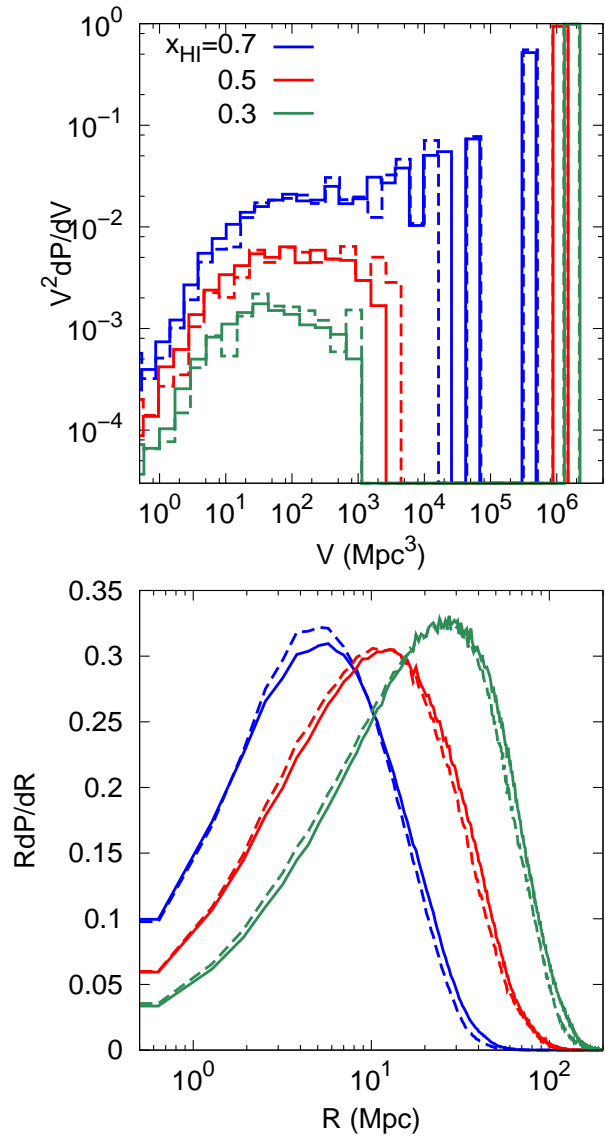}
   \caption{ Top panel: The volume distribution of the \HII ~bubbles calculated by the FOF method at three different stages of reionization in Model I. The left to right curves (with colours green, red and blue) correspond to redshift 9.5, 9 and 8.7 when the neutral fractions are 0.7, 0.5 and 0.3, respectively. The solid and dashed curves represent $\rm C^2$-{\sc ray} and {\sc grizzly} simulations respectively. Bottom panel: The PDF of the \HII ~bubble size calculated by the mean-free-path method. The line styles and colours have the same meaning as in the upper panel.}
   \label{image_p5_bubsize}
\end{center}
\end{figure}

\begin{figure*}
\begin{center}
\includegraphics[scale=0.75]{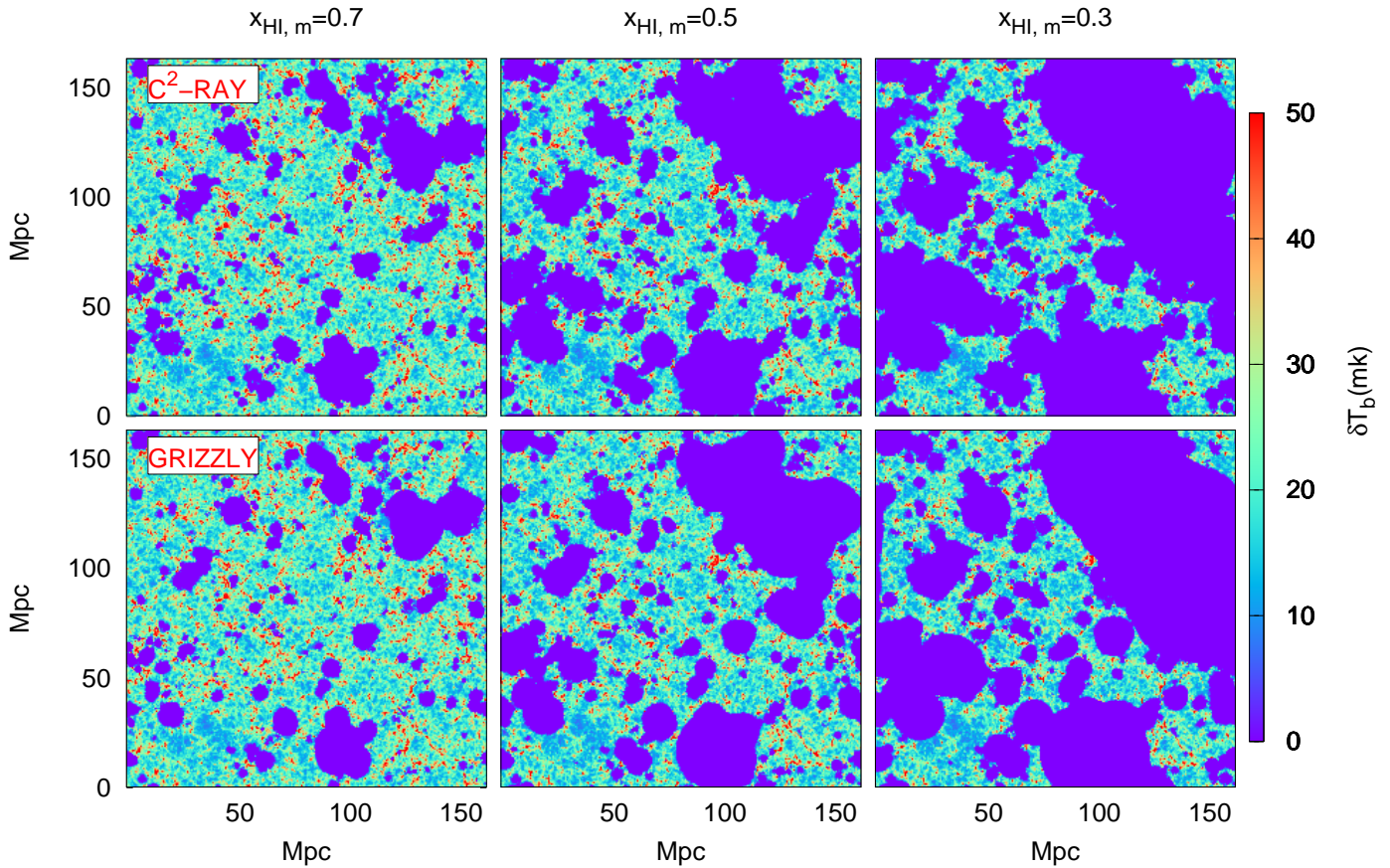}
\caption{The brightness temperature maps of the same slice as in Figure \ref{image_p5_mapxhizmid} for Model I. Top and bottom panels show the $\rm C^2$-{\sc ray} and {\sc grizzly} results, respectively. The left to right panels correspond to redshift 9.5, 9 and 8.7 with neutral fraction of 0.7, 0.5 and 0.3 respectively. }
\label{image_p5_maptb}
\end{center}
\end{figure*}

We next compare the \HII ~bubble size distribution (BSD) from these two simulations. 
BSDs can provide valuable information about the growth of the ionized regions \citep{friedrich2011topology}. However, due to the complex morphologies of the ionized regions, there is no unique way to define the sizes of the irregularly shaped \HII\ regions in simulation volume. Therefore different size determination methods have been developed and each address different aspects of the size distributions \citep{giri2017bubble}. In this study, we have use two such methods, namely the Friends-of-friends \citep[FOF;][]{2006MNRAS.369.1625I} and mean-free-path \citep[MFP;][]{2007ApJ...669..663M}.

First, we discuss our bubble size statistics from the FOF method. This algorithm considers two neighbouring cells with ionization fraction larger than 0.5 to be the part of the same ionized bubble and in the process finds the volume of each ionized region. This BSD algorithm focuses on connectivity. As shown in \citet{furlanetto2016reionization} the percolative nature of reionization leads to the development of one dominant connected region which contains most of the ionized volume.  This large connected region is known as the percolation cluster.
The single percolation cluster is invisible in the normalized histogram of the volumes (VdP/dV) as this quantity gives the number of ionized bubbles of volume V. Instead we plot $(V/V_\mathrm{ion})VdP/dV$ which shows the fraction of the ionized volume ($V_\mathrm{ion}$) contained in regions of volume $V$. Once the percolation cluster has formed it will have a value close to 1 in this quantity.

The top panel of Figure \ref{image_p5_bubsize} shows the FOF-BSDs from the two simulations at three different stages of reionization. The volume of the percolation cluster is identical in both the simulations. One can observe that the number and size distribution of smaller ionized regions also match quite well during the early and late stages of reionization.  However, we find that {\sc grizzly} produces a few larger bubbles during the intermediate stages. However, compared to the different semi-numerical models considered in \citet{2014arXiv1403.0941M} (as shown in Figure 5 of that paper), the FOF-BSD from {\sc grizzly} matches the $\rm C^2$-{\sc ray} results better.

The second BSD method that we use is MFP which is built on a Monte-Carlo inference of the sizes. This method selects random cells in the ionized regions and traces a ray in a random direction until it hits a neutral cell. In this case, we consider a cell to be ionized if its ionization fraction is larger than 0.5 and neutral if it is less than that. We repeat the ray tracing for enough number of times (in this case $10^7$ times) and record the lengths of the rays. This provides the estimate of the size distribution of the ionized regions, here expressed in radii instead of volumes.

The MFP-BSDs for {\sc grizzly} and $\rm C^2$-{\sc ray} show good agreement as can be seen in the bottom panel of Figure \ref{image_p5_bubsize}. The peaks of the curves at the different stages of reionization appear at the same radius which suggests that the most probable sizes of ionized regions in both simulations are identical. \citet{2014arXiv1403.0941M} did not calculate MFP-BSDs and therefore we cannot compare to their results in this case.   


\begin{figure}
\begin{center}
\includegraphics[scale=0.8]{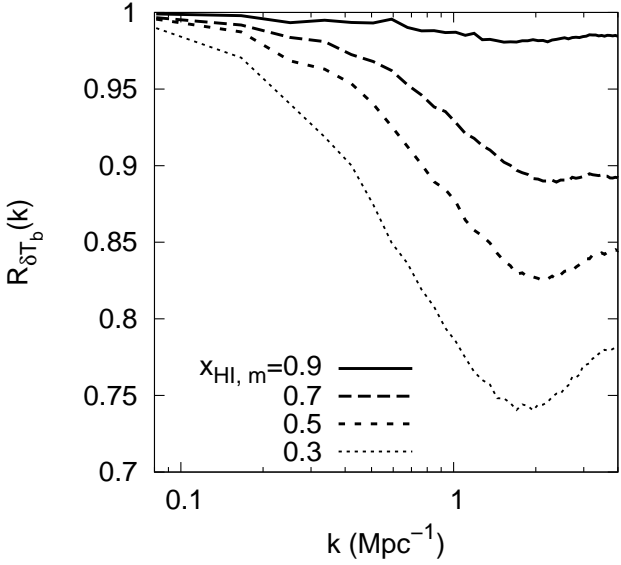}
    \caption{Scale dependence of the cross-correlation coefficient of the brightness temperature maps from $\rm C^2$-{\sc ray} and {\sc grizzly} for model I. Different curves represent redshifts 11.8, 9.5, 9 and 8.7 with neutral fraction 0.9, 0.7, 0.5 and 0.3 respectively.}
   \label{image_p5_crosstb}
\end{center}
\end{figure}

\subsubsection{Differential brightness temperature maps}
\label{21cmmaps}

The next quantity that we compare for the two simulations is the expected 21-cm signal. Figure \ref{image_p5_maptb} presents $\TB$ maps of the same slice at three different redshifts 9.5, 9 and 8.7 which correspond to the neutral fractions 0.7, 0.5 and 0.3 respectively. Visually the $\TB$ maps are quite similar. The Pearson-cross-correlation coefficients for these maps are 0.92, 0.9 and 0.9 respectively. However, the small-scale features are different due to the spherical nature of the bubbles around isolated sources in the 1D radiative transfer scheme. This was also seen in section \ref{sec_ion_map} when we considered the morphology of ionized regions. 

To check the level of similarity, we again use the scale-dependent cross-correlation between the two $\TB$ maps, see Figure \ref{image_p5_crosstb}. The large values of the cross-correlation coefficients $R_{\TB}$ suggest that the $\TB$ maps from these two schemes are very similar. However, we do find that the value of $R_{\TB}$ decreases at small scales which indicates mismatches between the $\TB$ maps at small scales. Still, $R_{\TB}$ is larger than 0.7 at scales such as $k \ge 1 ~\rm Mpc^{-1}$. This suggests that $\TB$ maps from {\sc grizzly} are more similar to the maps from $\rm C^2$-{\sc ray} than the maps from the semi-numerical models considered in \citet{2014arXiv1403.0941M}. For example, $R_{\TB}$ is less than 0.7 at $k \ge 1 ~\rm Mpc^{-1}$ for the semi-numerical models used in \citet{2014arXiv1403.0941M} while $R_{\TB} \sim 0.85$ for {\sc grizzly}.  We also find the $R_{\TB}$ value to be larger than the $R_{x_{\rm HI}}$ at scales $k \ge 1 ~\rm Mpc^{-1}$ (see Figure \ref{image_p5_crosstbxhi} ). This is due to the fact that the $\TB$ signal is dominated by the density fluctuation at small scales and these are identical between the two simulations.


\begin{figure}
\begin{center}
\includegraphics[scale=0.8]{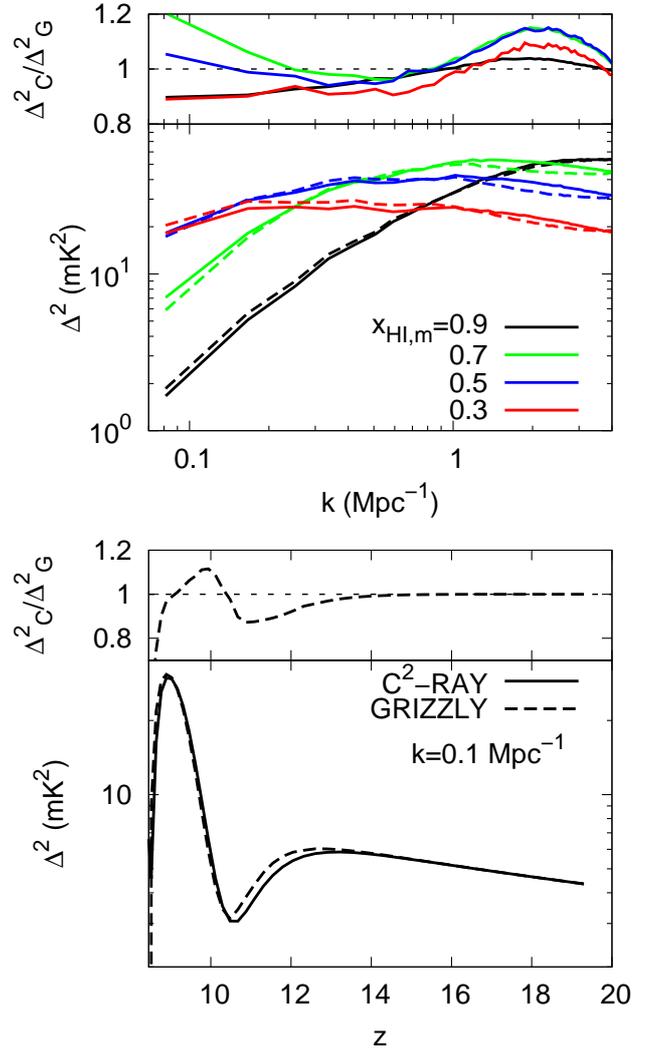}
    \caption{Top panel: Scale dependence of the spherically averaged dimensionless power spectrum of $\TB$ fluctuations at redshifts 11.8, 9.5, 9 and 8.7 with neutral fraction 0.9, 0.7, 0.5 and 0.3 respectively. The power spectrum corresponds to the Model I of the paper which considers ionizing photons from the massive halos only.  Bottom panel: Redshift evolution of the large-scale ($k=0.1 ~\rm Mpc^{-1}$) spherically averaged power spectrum estimated from $\rm C^2$-{\sc ray} and {\sc grizzly} which use  3D and 1D radiative transfer respectively. The ratios of the power spectrum from $\rm C^2$-{\sc ray} and {\sc grizzly} are also shown in the upper parts of the panels. ${\Delta}^2_{\rm C}$ and ${\Delta}^2_{\rm G}$ represent the dimensionless power spectrum from $\rm C^2$-{\sc ray} and {\sc grizzly} respectively.}
   \label{image_p5_psz}
\end{center}
\end{figure}


\subsubsection{Spherically averaged power spectrum of the signal}
\label{res_ps}
Next we compare the outputs of these two schemes in terms of the power spectrum of the $\TB$ fluctuations. The spherically averaged power spectrum $P(k)$ of the $\TB$ fluctuations is defined as
\begin{equation}
\langle \hat{\TB }(\rm{\bf k}) \hat{\TB}^{\star}(\bf{k'})\rangle = (2 \pi)^3 \delta_D(\bf{k - k'}) P(k),
\label{eqps}
\end{equation}
where $\hat{\TB}(\rm{\bf k})$ represents the Fourier transform of $\TB(\bf{x})$. Here, We present the results in terms of  the  spherically averaged dimensionless power spectrum ${\Delta}^2(k)=k^3 P(k)/2 {\pi}^2$. 

The top panel of Figure \ref{image_p5_psz} presents the scale dependence of ${\Delta}^2(k)$ estimated from these two simulations at different stages of reionization, together with the ratio of the two results. While the power spectrum is dominated by the density fluctuations at the early stages of reionization, it starts to deviate from the background dark matter density power spectrum as reionization progress. The power spectra from these two schemes agree with each other to within 10 percent for most stages and scales. Only at the largest scales for $x_\mathrm{HI}=0.7$ the difference exceeds 20 percent. The overall features of the power spectra are very similar between the two codes. 

The bottom panel of the figure compares the redshift evolution for the $k \sim 0.1 ~\rm Mpc^{-1}$ mode of the power spectrum. The ratio between the $\rm C^2$-{\sc ray} and {\sc grizzly} results starts to deviate from unity at redshift 14 and remains between 0.9 and 1.1 during the later stages of reionization once again indicating excellent agreement.
Compared to the semi-numerical methods from \citet{2014arXiv1403.0941M} the {\sc grizzly} results match the $\rm C^2$-{\sc ray} ones equally well, or better.


\subsection{Model II : Including low mass halos and thermal feedback}
\label{diff_source}
Finally we consider a more complex model for reionization and study the similarities of the outcomes of these two codes. While our fiducial model does not incorporate the contributions from LMACHs, these low-mass halos can provide a significant contribution to the overall ionizing photon budget. As explained in section \ref{source_rt}, in Model II we take the minimum mass of halos which contribute to reionization to be $10^8 ~\MSUN$ and we suppress star formation in halos of masses (below $10^9 ~\MSUN$) in ionized regions. This scenario is same as the $L1$ model in \citet{2012MNRAS.423.2222I}. The details can be found in Table \ref{tab1}. \citet{2014arXiv1403.0941M} did not study this model in their comparison between semi-numerical and numerical methods, so unlike for the results of Model I we are here not able to compare to semi-numerical methods.

\begin{figure}
\begin{center}
\includegraphics[scale=0.92]{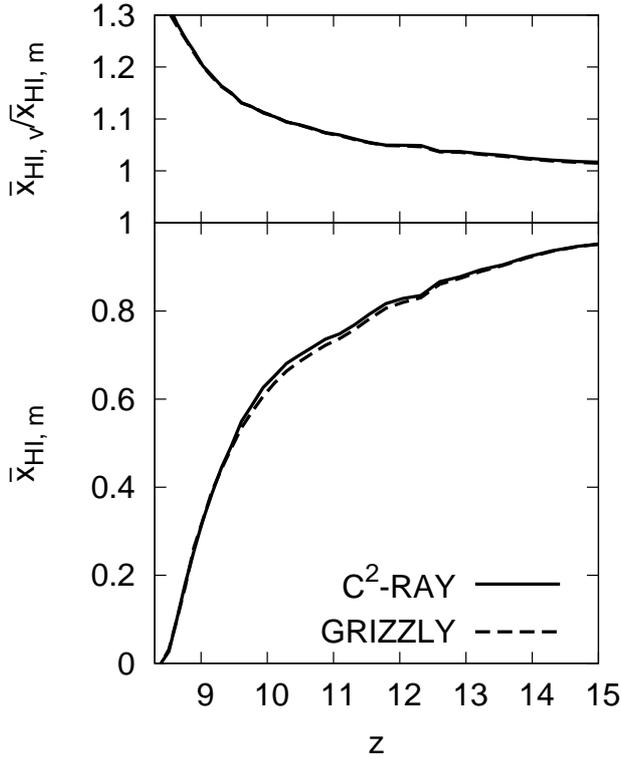}
    \caption{Bottom panel: Redshift evolution of the mass averaged neutral fraction from $\rm C^2$-{\sc ray} and {\sc grizzly} for the reionization scenario Model II which includes the contributions from low mass halos and considers thermal feedback. Top panel: Redshift evolution of the ratio of the volume and mass averaged neutral fraction from both the simulations. }
   \label{image_p5ionfrac_feed}
\end{center}
\end{figure}

\begin{figure}
\begin{center}
\includegraphics[scale=0.8]{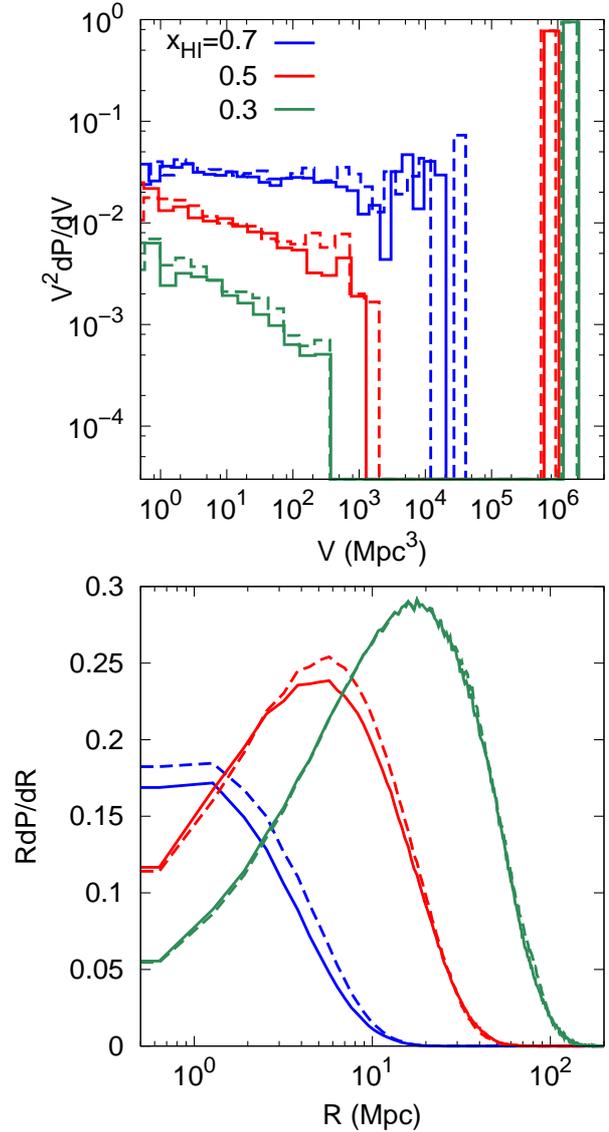}
    \caption{Top Panel: The volume distribution of the \HII\ bubbles as a function of volume at different stages of reionization for the two different simulations considered in this work. The left to right curves (with colours green, red and blue) correspond to ionization fraction 0.7, 0.5 and 0.3 at redshift 10.5, 9.5 and 8.9 respectively. The solid and dashed curves correspond to the  $\rm C^2$-{\sc ray} and {\sc grizzly} simulation respectively. Bottom panel: The PDF of the radius of the \HII ~bubbles as a function of the radius of the bubbles. The bubble size distribution in this panel is estimated using the mean-free path method. The reionization scenario considered here is Model II.}
   \label{image_p5_bubsize_feed}
\end{center}
\end{figure}

\begin{figure*}
\begin{center}
\includegraphics[scale=0.75]{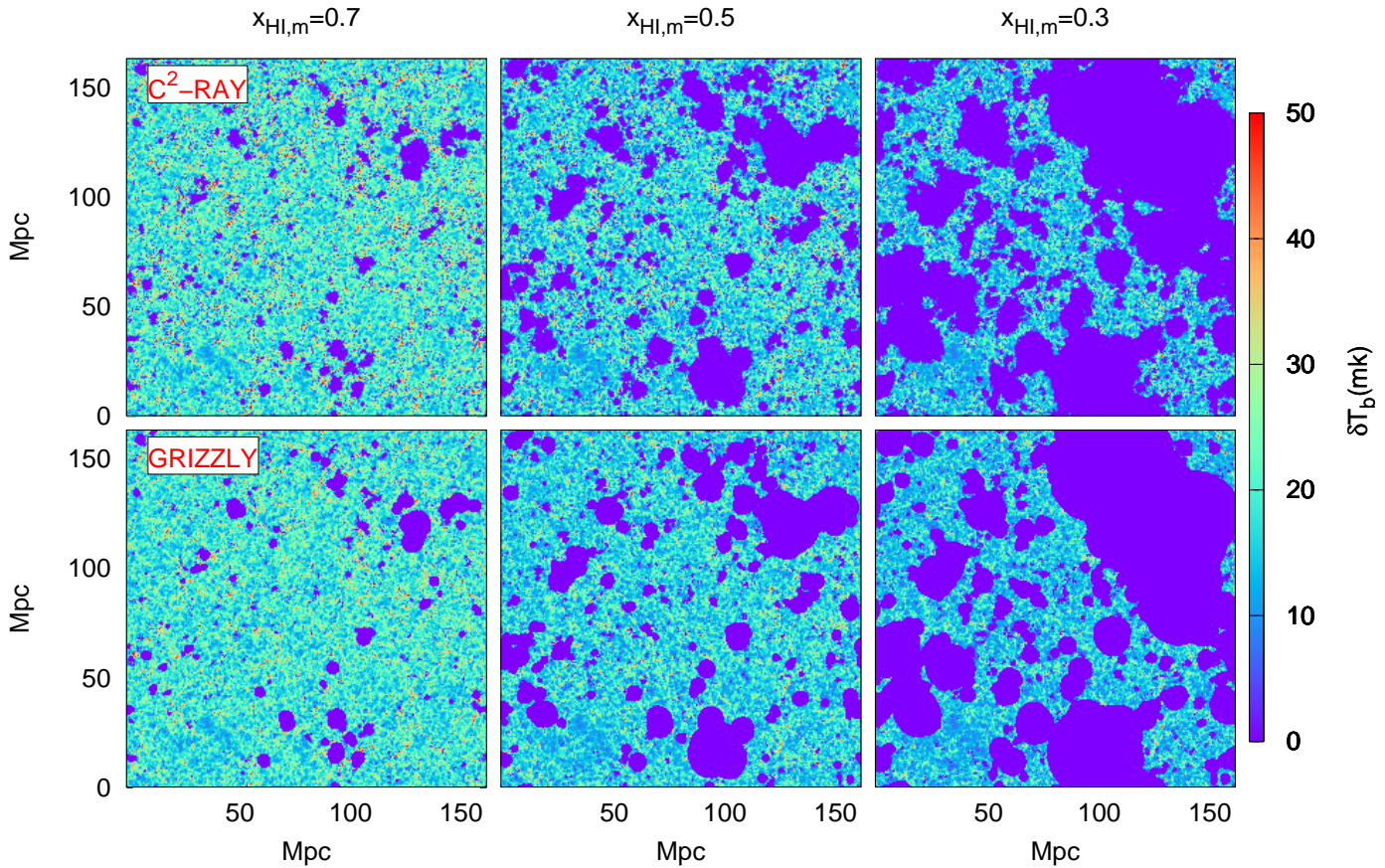}
    \caption{Top and bottom panels represent the $\TB$ maps corresponding to $\rm C^2$-{\sc ray} and {\sc grizzly} simulations respectively which are based on 3D and 1D radiative transfer schemes. The left to right panels represent redshift 10.5, 9.5 and 8.9 respectively, which correspond to neutral fraction 0.7, 0.5 and 0.3 respectively. The maps correspond to Model II which include the contributions from low-mass halos and consider thermal feedback.}
   \label{image_p5maptb_feed}
\end{center}
\end{figure*}

\begin{figure}
\begin{center}
\includegraphics[scale=0.8]{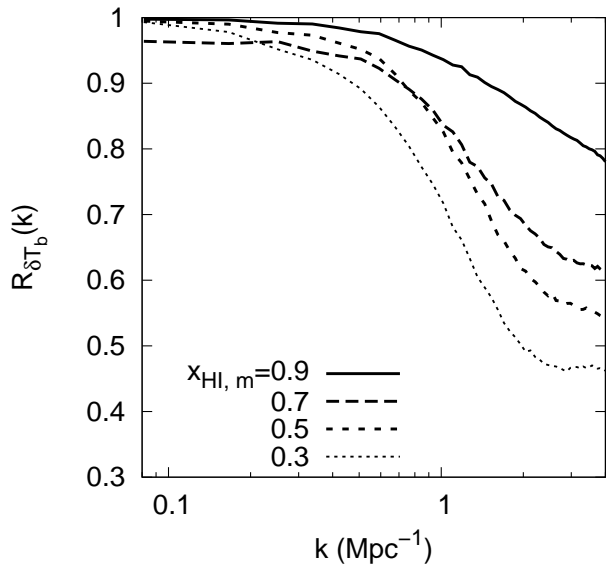}
    \caption{Scale dependence of the cross-correlation coefficient of the brightness temperature maps from $\rm C^2$-{\sc ray} and {\sc grizzly}  at different stages of reionization for Model II.}
   \label{image_p5_crosstb_feed}
\end{center}
\end{figure}

\begin{figure}
\centering
\includegraphics[scale=0.75]{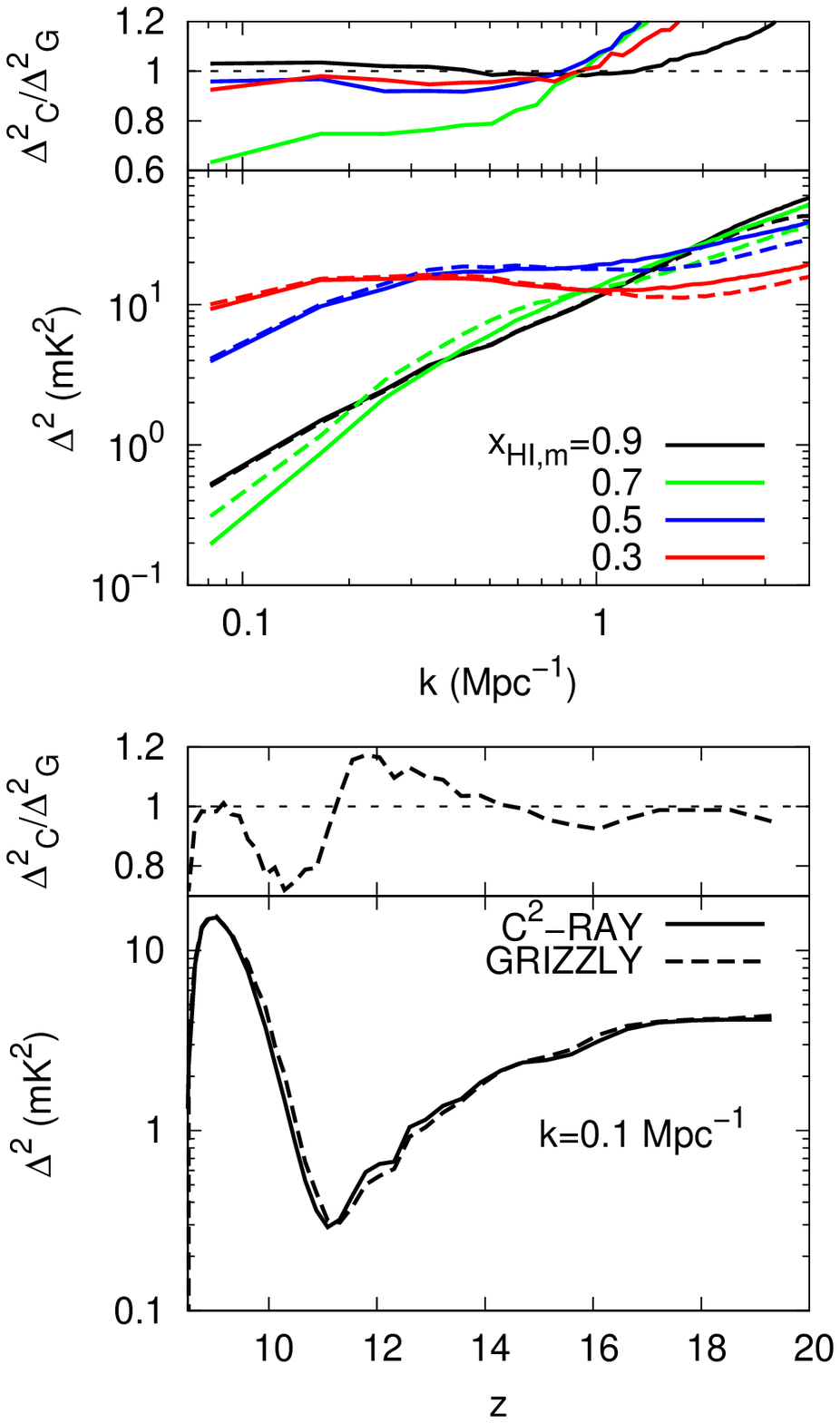}
\caption{Top panel: The spherically averaged power spectrum of the 21-cm signal for Model II as a function of scale at different stages of reionization. Different curves correspond to redshift 13.5, 10.5, 9.5 and 8.9 with neutral fraction 0.9, 0.7, 0.5 and 0.3 respectively.   Bottom panel: $\TB$ power spectrum for $k = 0.1 ~\rm Mpc^{-1}$ as a function of redshift. The solid and dashed curves in both the panels represent $\rm C^2$-{\sc ray} and {\sc grizzly} simulations respectively. Upper parts of the panels show the ratio of the power spectrum from  $\rm C^2$-{\sc ray} and {\sc grizzly} simulations. Here, ${\Delta}^2_{\rm C}$ and ${\Delta}^2_{\rm G}$ represent the dimensionless power spectrum from $\rm C^2$-{\sc ray} and {\sc grizzly} respectively.}
\label{image_p5_ps_feed}
\end{figure}

Although {\sc grizzly} has to apply simplifying approximations in order to include thermal feedback, the reionization history it generates is very similar to that from  $\rm C^2$-{\sc ray} as shown in the bottom panel of Figure \ref{image_p5ionfrac_feed}. The Thomson scattering optical depths for $\rm C^2$-{\sc ray} and {\sc grizzly} are 0.08 and 0.081 respectively. The redshift evolution of the ratio of the volume and mass averaged neutral fraction from these two simulations also matches very well as shown in the top panel of the figure. This suggests that the neutral fraction maps from these two simulations for this complex model are also very similar. We do not show these here but below will show the 21-cm images from which the morphology of the ionized regions can also be seen.

The FOF-BSDs obtained for Model II as shown in the top panel of Figure \ref{image_p5_bubsize_feed} have mostly very similar shapes. At the early stages (with neutral fraction 0.7), {\sc grizzly} is seen to produce more large ionized regions, although this difference is likely due to small size differences among the larger regions. However, this possible bias vanishes once the \HII\ regions start percolating during the middle and later stages of reionization. The distributions of the sizes of the \HII\ regions calculated using the MFP method for this scenario also display good agreement (as shown in the bottom panel of Figure \ref{image_p5_bubsize_feed}). We find that there is no prominent peak feature in the MFP-BSD curves during the early stages of reionization. This is due to the fact that the size of the ionized regions are very small and mostly fall within the resolution limit of the simulation.

The differential brightness temperature maps from these two simulations also visually very similar at different stages of reionization as shown in Figure \ref{image_p5maptb_feed}. The Pearson-cross-correlation coefficient calculated for the slices shown in Figure \ref{image_p5maptb_feed} are 0.7, 0.75 and 0.77 at reionization stages with neutral fraction 0.7, 0.5 and 0.3 respectively. The scale-dependent comparison of the $\TB$ cubes from these two simulations is also shown in terms of the cross-correlation coefficient at different stages of reionization in Figure \ref{image_p5_crosstb_feed}. One can see that the $\TB$ cubes from these two simulations are highly correlated at the large scales with $k \lesssim 0.5 ~\rm Mpc^{-1}$ (within 10 percent error), while the correlation drops at smaller scales.

We present the comparison between the spherically averaged power spectrum from these two simulations of Model II in Figure \ref{image_p5_ps_feed}. The redshift evolution of the spherically averaged power spectrum at scale $k\sim 0.1 ~\rm Mpc^{-1}$ are quite similar to each other as shown in the bottom panel of Figure  \ref{image_p5_ps_feed}. Although the power spectrum estimated from these two schemes are very similar, there are small differences too. For example, we find that the power spectra deviate from each other around redshift 10.5 when the neutral fraction is 0.7. Up to this time the LMACHs contribute substantially to reionization, so the differences are most likely connected to the way their suppression is implemented. Still, the differences are small and except at the very largest scales do not exceed 10 percent.

When comparing these results for Model II to the ones for Model I we see that {\sc grizzly} performs somewhat less well in the former case. However, the results are still very close to the numerical ones and definitely fall within the expected measurement errors from the observations.

Overall, we find that the results from these two simulations  agree well with each other even for this more complex scenario. This suggest that the 1D radiative transfer can be used for the complex scenarios as well without a substantial loss of accuracy. Specifically, the 3D and 1D radiative transfer schemes match quite well at scales $k \lesssim 1 $~Mpc$^{-1}$which are the prime target of the current radio interferometers. This suggests that 1D radiative transfer codes such as {\sc bears} or {\sc grizzly} can be efficiently used for predicting the expected signal for various source models as well as to find new observation strategies and possibly for parameter estimation.


\section{Summary and discussion}
\label{conc}
In this paper, we have compared the performance of the
1D radiative transfer code {\sc grizzly} to the 3D radiative transfer code $\rm C^2$-{\sc ray} in the context of a large scale reionization simulation. While $\rm C^2$-{\sc ray} simulates the process more accurately, by performing both the full 3D radiative transfer and following the evolution of the ionization fractions and the source in time, this accuracy comes at a considerable computational cost. The  {\sc grizzly} code simplifies the calculation by placing 
pre-generated 1D profiles of ionization fraction around the sources and using the halo growth history to approximate the 
evolutionary effects, which makes it approximately a factor $\sim 10^5$ faster.

We have used same sets of initial conditions, i.e.\  the same density fields, halo catalogues and source models for both simulations. We limit our comparison to the calculation of the ionization fraction. Beside comparing results around a single source, we consider two models to compare the outcomes of these simulations in detail. In Model I (also our fiducial model), we assume that the reionization is driven by massive halos with masses larger than $2.2\times 10^9 ~\MSUN$. This model was also used in \citet{2014arXiv1403.0941M} to evaluate the performance of semi-numerical reionization codes which rely on the excursion set formalism. We also consider a more complex model (Model II) which includes low mass halos (down to $10^8 ~\MSUN$) and suppression of star formation in halos with mass smaller than $10^9 ~\MSUN$ due to thermal feedback. For all the models considered in this paper, we have used the same SED which in this case is a blackbody spectrum. When calculating the the differential brightness temperature of the 21-cm signal we assume that $\TS \gg \TCMB$. Our findings from this comparison are the following. 

\begin{itemize}
\item When comparing the reionization history due to an isolated source, we observe that {\sc grizzly} underestimates the effect of recombinations. Since this is a cumulative effect, the impact is largest at the later stages. Still, we find that the differences between the neutral fractions estimated from the two methods for the isolated sources remain less than 10~percent.

\item For our fiducial simulation, the reionization histories in terms of the evolution of the mass averaged neutral fraction and ionization maps are very similar between both simulations. The cross-correlation coefficient of the neutral fraction maps remains close to unity (within 5-10 percent) at scales $k \lesssim 1 ~\rm Mpc^{-1}$ before the universe reaches 50 percent ionization and somewhat worse after this.

\item We have considered two different methods to characterize the size distribution of ionized regions, namely the friends-of-friends and mean-free path methods. For both methods the size distributions match very well between the $\rm C^2$-{\sc ray} and {\sc grizzly} results. However, visual inspection of the ionization fraction maps does show that {\sc grizzly} produces more spherically shaped regions compared to $\rm C^2$-{\sc ray}.

\item Visually, the differential brightness temperature maps from the two simulations look very similar. The cross-correlation coefficients at large scales, $k \lesssim 0.1 ~\rm Mpc^{-1}$ are close to unity. We find that the correlation coefficients at small scales are higher for the $\TB$ maps than for the ionization maps. This is due to the density fluctuations in the neutral medium.

\item The spherically averaged power spectra from the two simulations agree with each other within 10 percent.

\item For the more complex scenario Model II, we find that the differences between the {\sc grizzly} and $\rm C^2$-{\sc ray} are generally somewhat larger than for the simpler Model I but remain small. We conclude that {\sc grizzly} is capable of handling more complex source models where sources are not always ``on''.

\item The results from {\sc grizzly} are more similar to those from $\rm C^2$-{\sc ray} compared to the different semi-numerical simulations considered in  \citet{2014arXiv1403.0941M}. We find for example that the various cross-correlation coefficients are higher for {\sc grizzly}, especially for smaller scales. Furthermore, for the semi-numerical models the source efficiency was adapted at each redshift in order to reproduce the ionization history from $\rm C^2$-{\sc ray}, whereas {\sc grizzly} is able to produce an excellent match for the reionization history without varying the source efficiency parameter.

\end{itemize}

These results imply that the faster 1D radiative transfer codes such as {\sc bears} or {\sc grizzly} can be efficiently used for exploring parameter space, determining observational strategies or developing parameter estimation pipelines. Once the 1D profiles have been generated, these simulations are fast enough to be used for parameter estimation using techniques such as the Monte Carlo Markov Chains, machine learning, etc. However, the details of parameter estimation such as convergence time, accuracy, etc. are beyond the scope of this work and will be addressed in future works.

Although {\sc grizzly} is quite fast compared to the full radiative transfer simulations and generally reproduces the results well, our comparison also shows where improvements could be made. ($i$) {\sc grizzly} shows lower accuracy at small scales. We speculate that the accuracy at small scales can be improved by optimizing the technique {\sc grizzly} uses to deal with overlapping ionized regions. ($ii$) The comparison for a single source shows that {\sc grizzly} underestimates the effect of recombinations. This could be improved by taking into account the cosmological evolution of the density field over the life time of the source. ($iii$) The halo suppression model used in {\sc grizzly} is currently a very simple one. It may be possible to implement the evolution of relic \HII\ regions better and thereby to improve the accuracy of {\sc grizzly} for scenarios where source can turn ``off''.

Besides the neutral fraction maps, {\sc grizzly} can also provide the kinetic temperature and $\lya$ flux maps. Thus, it can be used to predict signal in presence of spin-temperature fluctuations. The detailed comparison of the results in presence of $\TS$ fluctuations with fully numerical simulations such as \citet{2017MNRAS.468.3785R} is beyond the scope of this paper and will be addressed in the future.

\section*{Acknowledgement}
The authors would like to thank Saleem Zaroubi and Adi Nusser for providing useful suggestions on this work. KKD would like to thank the University Grant Commission (UGC), India for support through UGC-faculty recharge scheme (UGC-FRP) vide ref.no. F.4-5(137-FRP)/2014(BSR). SM acknowledges the financial support from the European Research Council under ERC grant number 638743-FIRSTDAWN.

\bibliography{com_c2ray_grizzly}


\label{lastpage}
\end{document}